\documentstyle[aps,preprint]{revtex}
\begin{document}

\title {Effects of  multi-pion correlations on the source 
distribution in ultra-relativistic heavy-ion collisions}
\author{Q.H. Zhang}
\address{Institut f\"ur Theoretische Physik, Universit\"at Regensburg,
D-93040 Regensburg, Germany}
\vfill
\maketitle

\begin{abstract}
Multi-pion correlation effect on the source 
distribution is studied. It is shown that 
multi-pion Bose-Einstein correlation make the average 
radius of the pion source become smaller. The isospin effect on 
the pion multiplicity distribution and the source distribution is 
 also discussed.  
\end{abstract}

PACS number(s): 11.38 Mh, 11.30 Rd,25.75 Dw

Ultrarelativistic hadronic and nuclear collisions provide a unique 
 environment  to  create  dozens, and in some cases hundreds, of 
pions\cite{OPAL,NA35,E802}.  
To study the pion source distributions in these processes,  therefore, one   must  take 
into account  the effects of multi-pion 
Bose-Einstein (BE) 
correlations\cite{LL,WC84,Zajc87,Pratt93,PZ94,PGG90,APW93,ZC97,CGZ95,ZCG95,Zhang}. 
Among others, Lam and Lo \cite{LL} suggested that the 
larger isospin fluctuations  were due to the Bose nature of the emitted pions. 
Pratt \cite{Pratt93} suggested that if the number of bosons in a unit value of 
phase space is  large enough, bosons may condense into the 
same quantum state and a pion laser could be created. Considering isospin 
effect, Pratt and Zelevinsky \cite{PZ94} 
used the method  in Ref.\cite{Pratt93} to explain Centauro events\cite{LFH80,Andreev}. 
Zajc \cite{Zajc87} first  used Monte-Carlo method to analyse multi-pion 
Bose-Einstein correlation effects on two-pion interferometry.  
Detail derivation of multi-pion Bose-Einstein correlation can be found in Ref.\cite{CGZ95}.
The bosonic nature and isospin  of the 
pion should affect the single pion spectrum distribution in 
coordinate space.      However,  this  issue  
has not yet  been discussed in the  literature.        The purpose of this 
paper is  to analyse the effects of  multi-pion correlation and isospin 
 on the source distribution in coordinate space. 

   We begin with write down  the definition for 
the $n$-pion inclusive source distribution:
\begin{eqnarray}
P_{n}(x_{1},\cdot \cdot \cdot, x_{n})
=\langle j^*(x_1)\cdot\cdot\cdot j^*(x_n)
j(x_1)\cdot\cdot\cdot j(x_n)\rangle ,
\end{eqnarray}
which can be explained as the probability of observing $n$ pions at 
point $\{x_{i}, i=1,n\}$ all in the same $n$ pion event.
Here $j(x)$ is the current of pions,
 which can be expressed as\cite{PGG90,CGZ95,ZCG95}
\begin{equation}
j(x)=\int j(x',p)\exp\{ip(x-x')\}\gamma(x') d^4x'd^4p,
\end{equation} 
where $j(x',p)$ is the probability amplitude of finding a pion with momentum 
$p$, emitted by the emitter at $x'$. $\gamma(x')$ is a random phase 
factor which has been taken away  from $j(x',p)$. All emitters are uncorrelated 
in coordinate space if assuming:
\begin{equation}
\langle \gamma^*(x)\gamma(y)\rangle =\delta^4(x-y) .
\end{equation}
Taking the phase average and using  eq. (3), 
one can re-express the $n$-pion inclusive distribution (eq.(1)) as
\begin{eqnarray}
P_n(x_1,\cdot\cdot\cdot,x_n)&=&
\sum_{\sigma}\rho_{1,\sigma(1)}\cdot\cdot\cdot\rho_{n,\sigma(n)},
\end{eqnarray}
with 
\begin{eqnarray}
\rho_{i,j}=\rho(x_i,x_j)=<j^*(x_i)j(x_j)>,~~~\int \rho(x,x)d^4x=n_0 .
\end{eqnarray}
Here $\sigma(i)$ denotes the $i$th element of a permutation of the sequence
${1,2,3,\cdot \cdot \cdot, n}$, and the sum over $\sigma$  
runs  over all $n!$ permutations of this sequence.  $n_0$ is the mean 
pion multiplicity without multi-pion Bose-Einstein correlations. 

 Taking into account  the $n$-pion 
correlation effect, 
the normalized modified $i$-pion inclusive
distribution in $n$ pion events,
 $P_{i}^{n}(x_{1},\cdot \cdot \cdot, x_{i})$,
 can be written  as
\begin{equation}
P_{i}^{n}(x_{1},\cdot \cdot \cdot, x_i)=\frac{\int \prod_{j=i+1}^{n} d^4 x_{j}
P_{n}(x_{1},\cdot \cdot \cdot,x_{n})}{\int \prod_{j=1}^{n} d^4 x_{j}
P_{n}(x_{1},\cdot \cdot \cdot,x_{n})},
\end{equation}
which can be explained as the probability of finding $i$ pions at 
point $\{x_j, j=1,i\}$ 
in $n$ pion events. 

Now we define the function $G_{i}(x,y)$ as\cite{Pratt93,CGZ95}
\begin{equation}
G_{i}(x,y)=  \int \rho(x,x_{1}) d x_{1} \rho(x_{1},x_{2})
dx_{2} \cdot \cdot \cdot \rho(x_{i-2},x_{i-1})dx_{i-1}
\rho(x_{i-1},y).
\end{equation}

From the expression of $P_{n}(x_1,\cdot\cdot\cdot,x_n)$ (eq.(4)), 
the single pion inclusive distribution in coordinate space can be expressed as
\begin{eqnarray}
P_{1}^{n}(x)=
\frac{1}{n}\frac{1}{\omega(n)}\sum_{i=1}^{n}G_{i}(x,x)\cdot \omega(n-i),
\end{eqnarray}
with
\begin{eqnarray}
\omega(n)=\frac{1}{n!}\int P_n(x_1,\cdot\cdot \cdot,x_n) \prod_{k=1}^{n} d^4x_{k} .
\end{eqnarray}
Here $\omega(n)$ is the pion multiplicity distribution probability. 
Experimentally, one  usually mixes  all events  
to analyse the single-pion inclusive distribution.
Thus  the single-pion inclusive distribution reads 
\begin{equation}
P^{\phi}_1(x)
=\frac{\sum_{n=1}^{\infty}\omega(n)\cdot n
\cdot P_{1}^{n}(x)}
{<n> \sum_{n}\omega(n)},
\end{equation}
with 
\begin{equation}
<n>=\frac{\sum_{n=1}^{\infty}\omega(n)\cdot n}
 {\sum_{n}\omega(n)} .
\end{equation}
Using eq.(8),  one can writes down the 
single-pion inclusive distribution 
\begin{equation}
P_1^{\phi}(x)=\frac{1}{<n>}\sum_{i=1}^{\infty} G_{i}(x,x) .
\end{equation}
In the following, we  will employ a simple model to analyse 
the multi-pion correlation effects on the source 
distribution. We assume that 
\begin{equation}
\rho(x,y)=n_0\cdot (\frac{1}{\pi R_1^2})^{3/2}
\exp(-(\vec{x}+\vec{y})^2/4R_1^2)
\exp(-(\vec{x}-\vec{y})^2/4R_2^2)
\sqrt{\delta(x_0)\delta(y_0)} .
\end{equation}
Here $R_1$ and $R_2$  are the parameters  that represents the radius of 
the chaotic source and the correlation length of pions 
respectively.  $x=(x_0,\vec x) $ are the pion's 
four dimensional coordinate.
Then,  $G_n(x,y)$ can be expressed as
\begin{equation}
G_n(x,y)=n_0^n\cdot\sqrt{\delta(x_0)\delta(y_0)}\cdot \alpha_n\cdot 
\exp\{-a_n\cdot(\vec x^2+\vec y^2) + g_n\cdot \vec x\cdot 
\vec y\},
\end{equation}
where
\begin{eqnarray}
a_{n+1}=a_1-\frac{g_1^2}{4b_n},g_{n+1}=\frac{g_1\cdot g_n}{2b_n},
b_n=a_n+a_1,
\end{eqnarray}
and
\begin{equation}
\alpha_{n+1}=\alpha_n\cdot(\frac{1}{b_n})^{3/2} \cdot(\frac{1}{R_1^2})^{3/2},
\end{equation}
with
\begin{equation}
a_1=\frac{1}{4R_1^2}+\frac{1}{4R_2^2},g_1=\frac{1}{2R_2^2}-
\frac{1}{2R_1^2},\alpha_1=(\frac{1}{\pi R_1^2})^{3/2} .
\end{equation}

From the above formula,  we obtain 
 the pion source distribution in $n$ pion events ($P_1^n(\vec x)=\int P_1^n(x) dx_0$), 
which we show in Fig. 1. 
 Evidently,  due to the multi-pion Bose-Einstein correlation 
effects, the pions are concentrated in the same 
state. The mean radius of the source becomes smaller. The 
larger the pion multiplicity, 
the larger the BE correlation effects on the source 
distribution.   Multi-pion correlation effects on the 
source distribution ($P_1^{\phi}(\vec x)=\int P_1^{\phi}(x)dx_0$) is
 shown in fig.2, 
where we mix the 
pion multiplicity. It is clear that as the mean pion multiplicity becomes 
larger, the multi-pion correlation effects on the source 
distribution become larger. 

 The pion state with isospin can be written as\cite{APW93,HS71} 
\begin{equation}
|\phi>=\exp(\int \vec{j}(x)\cdot \vec{a}^+(x))|0>,
\end{equation}
where $\vec{j}$ and $\vec{a}$ are the vectors in isospin space which can be expressed as
\begin{equation}
\vec{j}(x)=j(x)[\frac{1}{\sqrt{2}}\sin\theta e^{-i\phi},
\cos\theta,
-\frac{1}{\sqrt{2}}\sin(\theta)e^{i\phi}],
\end{equation}
and
\begin{equation}
\vec{a}^+(x)=[a_{\pi^-}^+(x)
,a_{\pi^0}^+
,a_{\pi^+}^+].
\end{equation}
Then the state with total isospin $I$ and $z$ component of isospin $I_z$ is 
\begin{equation}
|\phi,I,I_z>=\int 
\sin\theta d\theta d\phi Y_{I,I_z}^*(\theta,\phi)
exp(\int \vec{j}(x)\cdot \vec{a}^+(x))|0>.
\end{equation}

$Y_{I,I_z}(\theta,\phi)$ is the spherical harmonic 
of angular momentum $I$ and $z$ component $I_z$. From the above 
formula we can calculate isospin effect on pion probability and 
pion source distribution.   For the sake of simplicity,
 we will only consider  in the following the 
$I=0,I_z=0$ case.  One can  easily check that $|\phi,0,0>$ can be  
expanded in Fock space as 
\begin{equation}
|\phi,0,0>=\sum_{n_{\pi^0}} \sum_{n_c/2}\frac{1}{2^{n_c/2}}B[\frac{1}{2}+
\frac{n_{\pi^0}}{2},\frac{n_c}{2}+1]|n_{\pi^0}>|\frac{n_c}{2}>
|\frac{n_c}{2}>,
\end{equation}
with 
\begin{equation}
|n_{\pi^0,\pi^+,\pi^-}>=\frac{(\int j(x)a^+_{\pi^0,\pi^+,\pi^-}d^4x)^n}
{n!}|0>.
\end{equation}
Here $|n_{\pi}>$ denote $n$ pion state and $B(x,y)$ is the Beta function.  
$n_c=n_{\pi^+}+n_{\pi^-}$ is the total number of charged pions. 
Due to the isospin conservation, 
we have $n_{\pi^+}=n_{\pi^-}=\frac{n_c}{2}$ and 
$n_{\pi^0}$ must be even.  At this stage we 
 can calculate $\pi^0$ and $\pi^{+,-}$ probability distribution
according to 
\begin{equation}
P(n_{\pi^0})=\frac{\sum_{n_c}\omega(n_{\pi^0})\frac{1}{2^{n_c}}
B^2(\frac{n_{\pi^0}+1}{2},
\frac{n_c}{2}+1)\omega^2(\frac{n_c}{2})}
{\sum_{n_{\pi^0}}\sum_{n_c}\omega(n_{\pi^0})\frac{1}{2^{n_c}}
B^2(\frac{n_{\pi^0}+1}{2},
\frac{n_c}{2}+1)\omega^2(\frac{n_c}{2})},
\end{equation}
and
\begin{equation}
P(n_{\pi^+}=n_{\pi^-}=\frac{n_c}{2})=
\frac{\sum_{n_{\pi^0}}\omega(n_{\pi^0})\frac{1}{2^{n_c}}
B^2(\frac{n_{\pi^0}+1}{2},
\frac{n_c}{2}+1)\omega^2(\frac{n_c}{2})}
{\sum_{n_{\pi^0}}\sum_{n_c}\omega(n_{\pi^0})\frac{1}{2^{n_c}}
B^2(\frac{n_{\pi^0}+1}{2},
\frac{n_c}{2}+1)\omega^2(\frac{n_c}{2})}.
\end{equation}
The $\pi^0$ and $\pi^{+,-}$ probability distributions 
are  shown in Fig.3. For comparison, the pion 
probability without isospin conservation is also 
shown as solid line in Fig.3.  We  notice 
that due to isospin effect, the $\pi^0$ and $\pi^{+,-}$ 
probability distributions are quite different 
from each other now.  This phenomena have ever  been noticed 
in Ref.\cite{PZ94}. But 
the method given here enable us to calculated the 
isospin effect for any isospin state, 
 not limited to iso-singlet state as discussed in Ref.\cite{PZ94}. 
From the 
$\pi^0$ and $\pi^{+,-}$ probability distributions, we 
acquire  the mean multiplicities  
$<n_{\pi^0}>$ and $<n_{\pi^{+,-}}>$ 
\begin{equation}
<n_{\pi^0}>=
\frac{\sum_{n_{\pi^0}}\sum_{n_c}\frac{1}{2^{n_c}}B^2(\frac{n_{\pi^0}}{2}+\frac{1}{2},
\frac{n_c}{2}+1)\omega^2(\frac{n_c}{2}) n_{\pi^0}\omega(n_{\pi^0})}
{\sum_{n_{\pi^0}}\sum_{n_{c}}\frac{1}{2^{n_c}}B^2(\frac{n_0}{2}+\frac{1}{2},
\frac{n_c}{2}+1)\omega^2(\frac{n_c}{2})\omega(n_{\pi^0})},
\end{equation}
and	
\begin{equation}
<n_{\pi^{+,-}}>=
\frac{\sum_{n_{\pi^0}}\sum_{n_c}\frac{1}{2^{n_c}}B^2(\frac{n_{\pi^0}}{2}+\frac{1}{2},
\frac{n_c}{2}+1)\omega^2(\frac{n_c}{2}) n_{\pi^{+,-}}\omega(n_{\pi^0})}
{\sum_{n_{\pi^0}}\sum_{n_{c}}\frac{1}{2^{n_c}}B^2(\frac{n_0}{2}+\frac{1}{2},
\frac{n_c}{2}+1)\omega^2(\frac{n_c}{2})\omega(n_{\pi^0})}.
\end{equation}

It is well known that  in the case of 
the  isospin singlet, if  the  multi-particle BE correlation 
in coordinate space is neglected ($R_2=0 fm$),
 there would be  $<n_{\pi^0}>=<n_{\pi^{+}}>=<n_{\pi^-}>$\cite{HS71}. 
As the multi-particle BE correlation is included
 in coordinate space, however,  such a simple relation
 does not hold for  the isosinglet any more. 
As a example,  to make a quantitative sense, we set 
 $R_1=5 fm, R_2= 0.8 fm $ and $n_0=20$ and obtain  
$n_{\pi^0}=2.86$ and $n_{\pi^+}=n_{\pi^-}=9.38$. 

The $\pi^0$ and $\pi^{+,-}$ source distribution reads  
\begin{eqnarray}
P_{\pi^0}(x)&=&\frac{<\phi,0,0|a_{\pi^0}^+(x)a_{\pi^0}(x)|\phi,0,0>}
{<\phi,0,0|\phi,0,0>< n_{\pi^0} >}
\nonumber\\
&=&\frac{1}{<n_{\pi^0}>}
\frac{\sum_{n_{\pi^0}}\sum_{n_c}\frac{1}{2^{n_c}}B^2(\frac{n_{\pi^0}}{2}+\frac{1}{2},
\frac{n_c}{2}+1)\omega^2(\frac{n_c}{2})\sum_{i=1}^{n_{\pi^0}}G_i(x,x)\omega(n_{\pi^0}-i)}
{\sum_{n_{\pi^0}}\sum_{n_{c}}\frac{1}{2^{n_c}}B^2(\frac{n_{\pi^0}}{2}+\frac{1}{2},
\frac{n_c}{2}+1)\omega^2(\frac{n_c}{2})\omega(n_{\pi^0})},
\end{eqnarray}
and 
\begin{eqnarray}
P_{\pi^{+,-}}(x)&=&\frac{<\phi,0,0|a_{\pi^{+,-}}^+(x)a_{\pi^{+,-}}(x)|\phi,0,0>}
{<\phi,0,0|\phi,0,0>< n_{\pi^{+,-}} >}
\nonumber\\
&=&\frac{1}{<n_{\pi^{+,-}}>}
\frac{\sum_{n_{\pi^0}}\sum_{n_c}\frac{1}{2^{n_c}}B^2(\frac{n_{\pi^0}}{2}+\frac{1}{2},
\frac{n_c}{2}+1)\omega(n_{\pi^0})\omega(\frac{n_c}{2})
\sum_{i=1}^{n_{\pi^{+,-}}}G_i(x,x)\omega(\frac{n_{c}}{2}-i)}
{\sum_{n_{\pi^0}}\sum_{n_{c}}\frac{1}{2^{n_c}}B^2(\frac{n_{\pi^0}}{2}+\frac{1}{2},
\frac{n_c}{2}+1)\omega^2(\frac{n_c}{2})\omega(n_{\pi^0})}.
\end{eqnarray}

The $\pi^0$ and $\pi^{+,-}$ inclusive distributions are 
 shown in Fig.4.  For comparison, we also give  the 
$\pi$ inclusive distribution without isospin effect in the same figure. 
 Notice that there are  a difference among 
$\pi^0$,$\pi^{+,-}$ and $\pi$ inclusive distribution caused by 
multi-particle BE correlations in coordinate space. 

In this paper, we  have  not considered   the 
energy constraint effects on the source distribution as 
done in Ref. \cite{ZCG95}. Also the source model present here 
is not a realistic model. 
As  stressed in Ref.  \cite{Zajc87,PZ94}, for a more or less real model,
the amount of  calculational work  will increase  astronomically. 
To our best knowledge,  there is so far no  method
 which enables  us to calculate quickly multi-particle 
BE correlation in the above process. 
But we think that for  the  purpose of illustrating 
the general  features of multi-pion correlation effects 
on the source distribution, our toy model is  good enough. 
In Ref. \cite{FF91}, Fowler et al. found that the 
restriction on the multiplicity have great influence on BE correlation. 
 We  did not address this issue, simply because   our prime 
purpose on the paper is to discuss 
the multi-pion BE correlation and isospin effects on the 
source distribution in coordinate space, 
 which has not yet been  discussed in the  
previous publication. We  leave the systematical 
  discussion of all these  factors on the  source 
distribution  for the future publication.

  In conclusion, the multi-pion Bose-Einstein 
correlation effects on the pion source 
distribution  has been discussed.  We showed  that 
multi-pion correlation make the average 
radius of the source become smaller. 
The larger the pion multiplicity,  the larger the multi-pion correlation 
effects on the source distribution. Isospin effects 
on pion probability and 
pion source distribution  are  also discussed.  We observed that 
multi-pion correlations distort the relation 
$<n_{\pi^0}>=<n_{\pi^+}>=<n_{\pi^-}>$ 
which exists in isospin-singlet state without 
multi-particle BE correlation 
in coordinate space.

\begin{center}
{\bf Acknowledgement}
\end{center}
The author  thanks  Dr. Y. Pang 
for helpful discussions and to Dr. W. Lu for reading the manuscript.  
This work was partly supported by 
the Alexander von Humboldt foundation in Germany.

\begin{center}
{\bf Figure Captions}
\end{center}

\end {document}
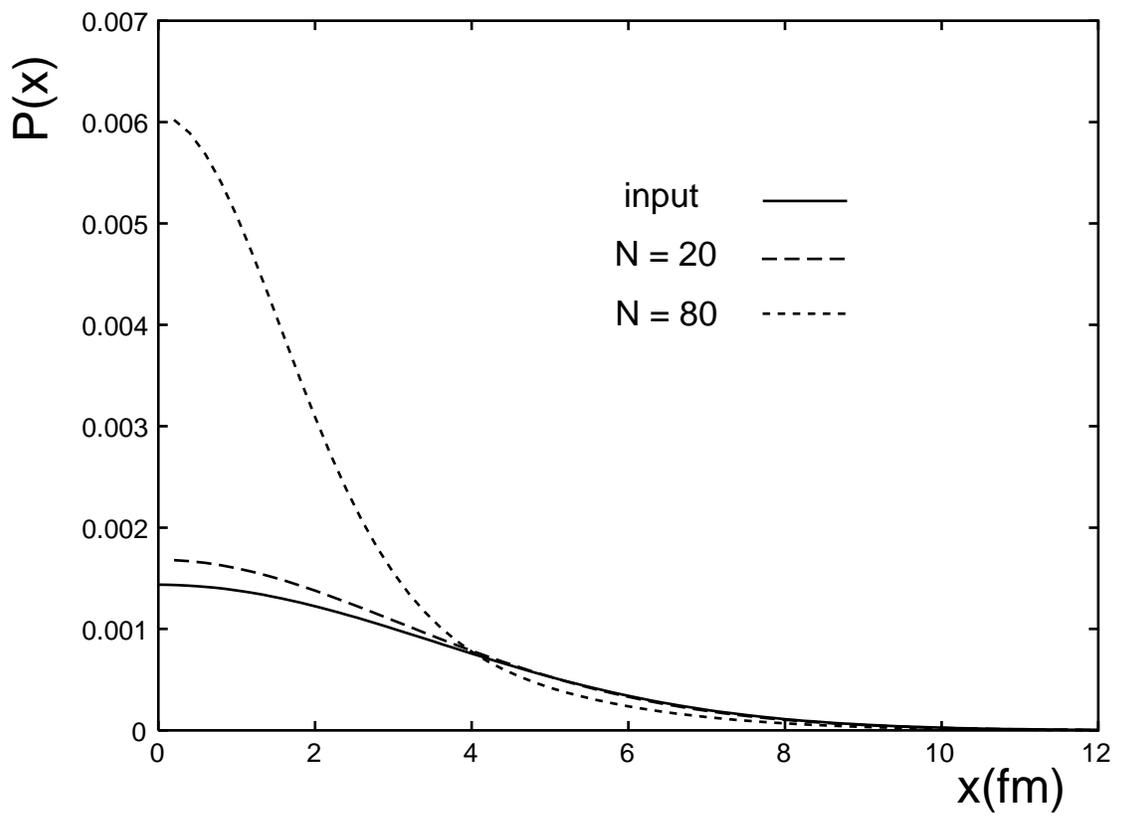
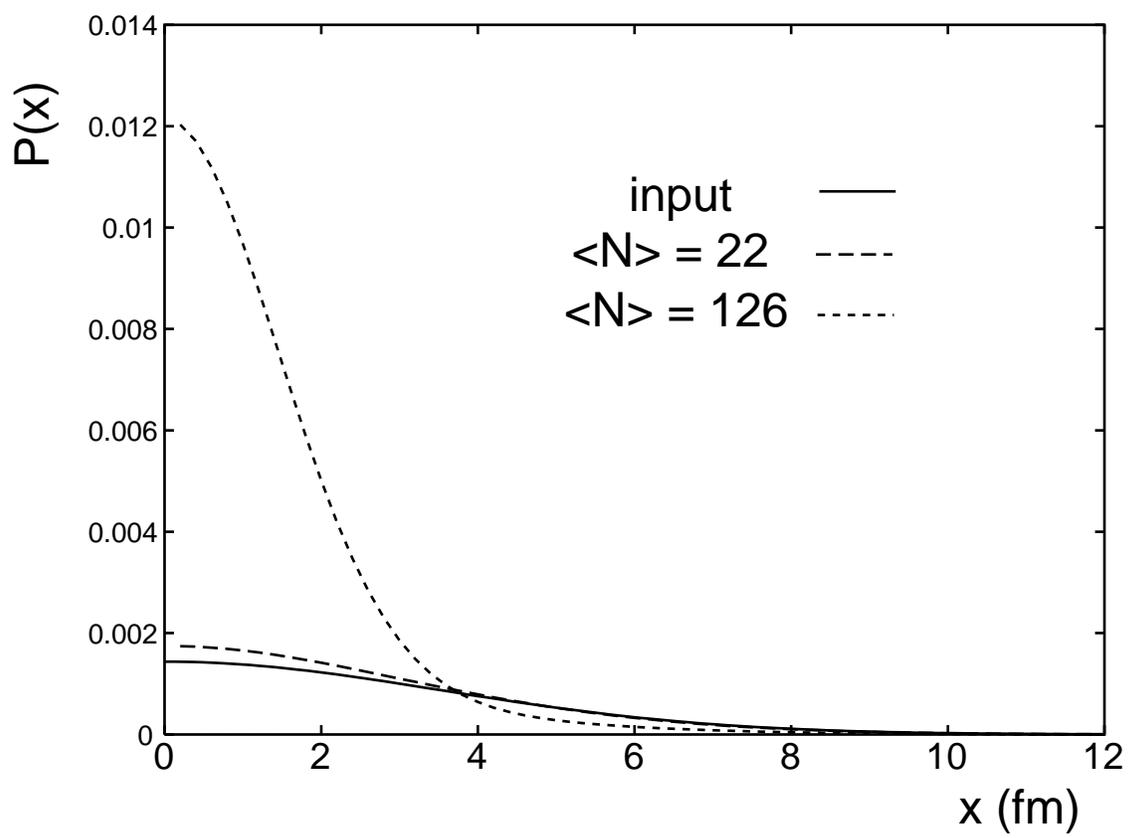
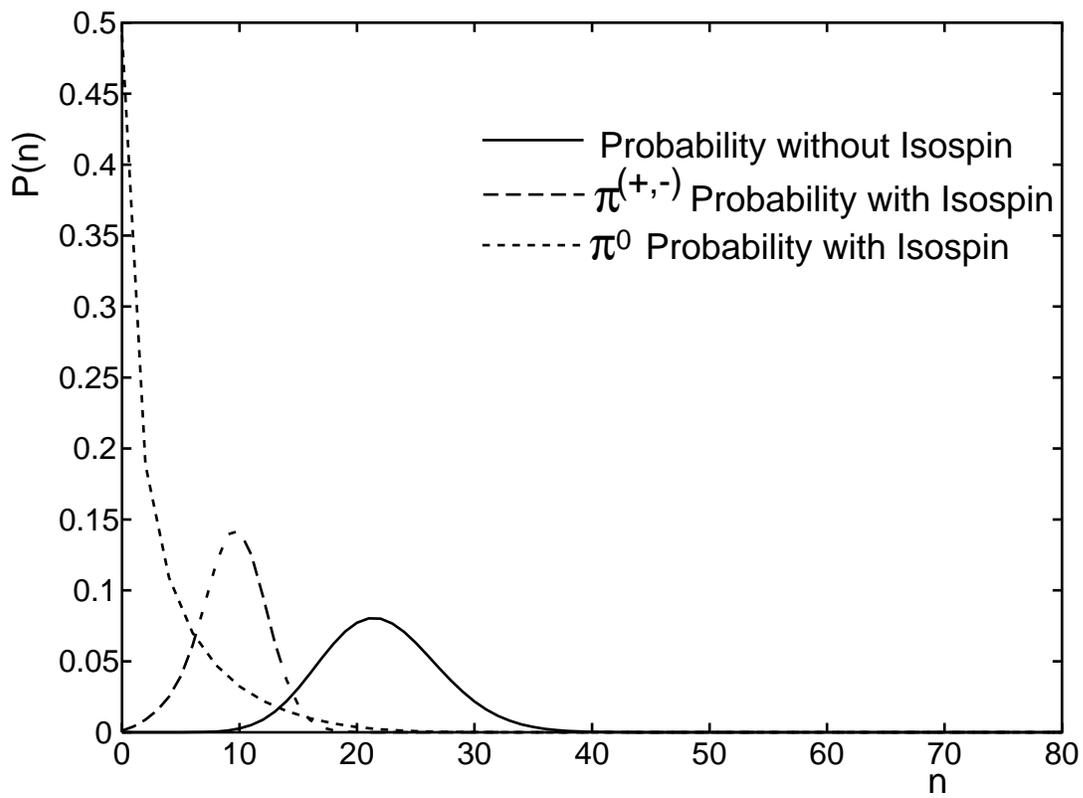
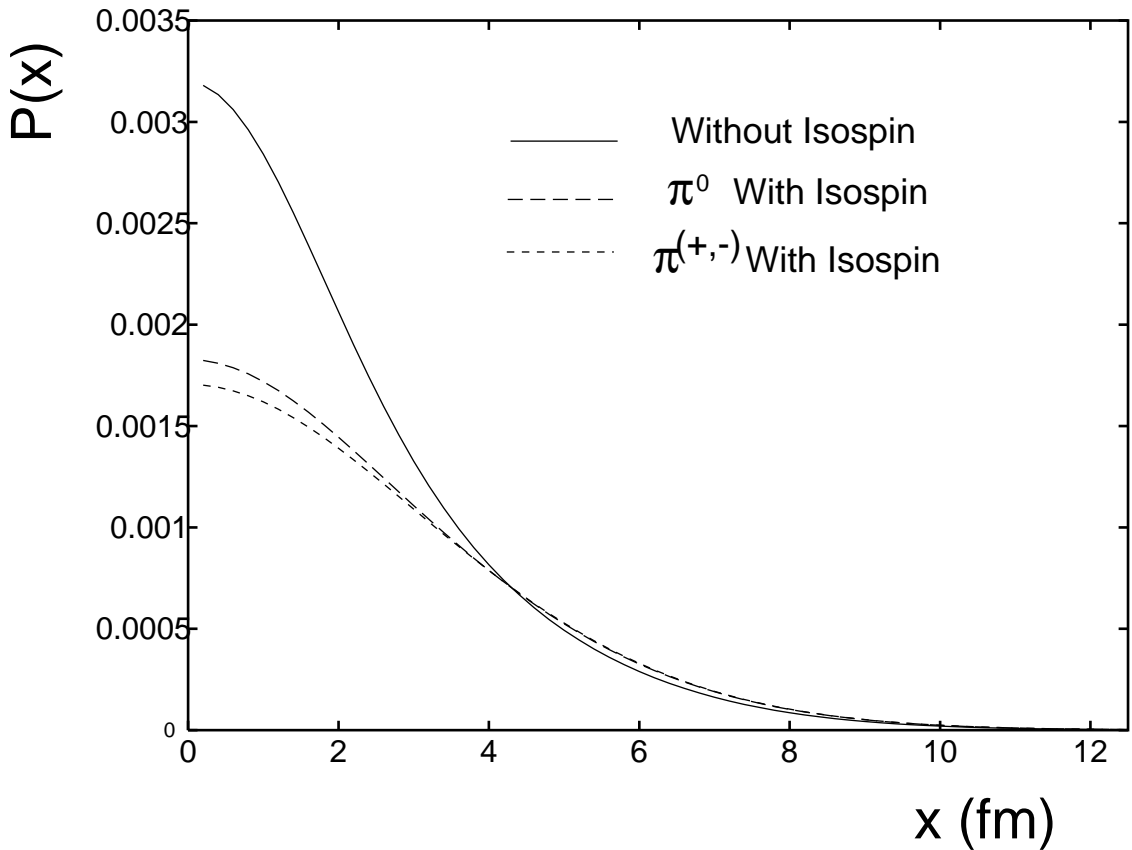
